\documentclass[9pt,conference]{IEEEtran}
\IEEEoverridecommandlockouts
\usepackage{cite}
\usepackage[table]{xcolor} 
\usepackage{amsmath,amssymb,amsfonts}
\usepackage{algorithmic}
\usepackage{graphicx}
\usepackage{textcomp}
\usepackage{multirow}
\usepackage{tablefootnote}
\usepackage{adjustbox}
\usepackage{booktabs}
\usepackage{svg}
\usepackage{makecell}

\definecolor{lightgray}{gray}{0.90} 
\definecolor{headergray}{gray}{0.85} 
\definecolor{mygreen}{RGB}{40,130,40} 
\definecolor{myred}{RGB}{190,50,50}   

\def\BibTeX{{\rm B\kern-.05em{\sc i\kern-.025em b}\kern-.08em
    T\kern-.1667em\lower.7ex\hbox{E}\kern-.125emX}}
\begin{document}
\title{ A correlation-permutation approach for speech-music encoders model merging
}

\author{%
  \begin{tabular}{c c c} 
    
    \parbox[t]{.3\textwidth}{\centering
      \textbf{Fabian Ritter-Gutierrez} \\
      \textit{Nanyang Technological University}\\
      Singapore \\
      s220064@e.ntu.edu.sg
    } &
    \parbox[t]{.3\textwidth}{\centering
      \textbf{Yi-Cheng Lin} \\
      \textit{National Taiwan University}\\
      Taiwan \\
      f12942075@ntu.edu.tw
    } &
    \parbox[t]{.3\textwidth}{\centering
      \textbf{Jeremy H.M Wong} \\
      \textit{Institute for Infocomm Research}\\
      Singapore \\
      jeremy\_wong@i2r.a-star.edu.sg
    } \\
    
    \\[2ex] 
    \parbox[t]{.3\textwidth}{\centering
      \textbf{Hung-yi Lee} \\
      \textit{National Taiwan University}\\
      Taiwan \\
      tlkagkb93901106@gmail.com
    } &
    \parbox[t]{.3\textwidth}{\centering
      \textbf{Eng Siong Chng} \\
      \textit{Nanyang Technological University}\\
      Singapore \\
      aseschng@ntu.edu.sg
    } &
    \parbox[t]{.3\textwidth}{\centering
      \textbf{Nancy F. Chen} \\
      \textit{Institute for Infocomm Research}\\
      Singapore\\
      nfychen@i2r.a-star.edu.sg
    }
  \end{tabular}
}


\maketitle

\begin{abstract}
Creating a unified speech and music model requires expensive pre-training. Model merging can instead create an unified audio model with minimal computational expense. However, direct merging is challenging when the models are not aligned in the weight space. Motivated by Git Re-Basin, we introduce a correlation-permutation approach that aligns a music encoder's internal layers with a speech encoder. We extend previous work to the case of merging transformer layers. The method computes a permutation matrix that maximizes the model's features-wise cross-correlations layer by layer, enabling effective fusion of these otherwise disjoint models. The merged model retains speech capabilities through this method while significantly enhancing music performance, achieving an improvement of 14.83 points in average score compared to linear interpolation model merging. This work allows the creation of unified audio models from independently trained encoders.
\end{abstract}

\begin{IEEEkeywords}
Model merging, self-supervised learning, audio foundation models
\end{IEEEkeywords}

\section{Introduction}

Speech and music are two fundamental domains of audio, each with unique characteristics and complexities. Significant advancements have been made in self-supervised learning (SSL) for both areas, with encoders like HuBERT \cite{hubert} demonstrating exceptional performance in speech representation and MERT \cite{mert} tailored for music understanding. These models have proven effective in capturing domain-specific features through large-scale pre-training. However, current SSL models often treat speech and music in isolation \cite{ritter_distill_speech_music}. While this benefits domain-specific tasks, the increasing demand for unified audio understanding in applications such as audio large language models necessitates a single, general model capable of processing both domains \cite{dynamicsuperbv2}. Training such a comprehensive model is computationally expensive \cite{compute_budget}.

An alternative to costly pre-training is model merging, which involves combining parameters from different models to create a single model that inherits their capabilities \cite{task_arithmetic, tiesmerging, dare}. Model merging has been successfully explored in the speech technology domain for supervised fine-tuning scenarios \cite{task_arithmetic_speech_translation,task_arithmetic_synthetic_speech,task_arithmetic_tts,selective_attention_merging_asr,ritter_distill_speech_music}. Nonetheless, there is a key limitation in all previous work on model merging for speech, which is the assumption of starting from a commonly initialized model. Such limitation arises from the requirement that for effective merging, models' weight spaces must be similar enough to ensure their weights are linearly mode connected \cite{permutation_linear_interpolation}. This assumption is feasible in the case of fine-tuning the same model, as their weight space does not deviate much from the initialized models. Nonetheless, this is not the case for the situation where 2 completely independently-trained models are attempted to be merged.

Merging independently trained models, however, is not a trivial task. Naive approaches like simple weight averaging often lead to a significant degradation in performance, as shown in Table \ref{tab:main_results}. This is because models trained independently, even with the same architecture, tend to converge to different, symmetrically equivalent minima in the loss landscape \cite{permutation_linear_interpolation,git-rebasin,zipit}. Direct interpolation of weights in such cases can place the merged model in a suboptimal region, leading to a collapse in performance. As highlighted by the ZipIt \cite{zipit} framework, these minima are often connected by low-loss paths only when their internal representations, such as neurons or channels, are appropriately permuted and aligned.

The present paper addresses the problem of merging independently trained models by maximizing channel-wise cross-correlations between the feature activations of the models to find the best channel alignments between them; such an approach allows the merging of models that were trained with their own loss functions and pre-training data. The foundational works of Git-Re Basin \cite{git-rebasin} and Zipit \cite{zipit} demonstrated the feasibility of merging independently trained models for MLP-based and CNN-based architectures on the computer vision domain. 
A subsequent work \cite{merging_text_transformers} provided insights on how transformer \cite{transformer} layers for the text domain should be merged when following a permutation framework. While they provided evaluations on loss barriers of transformer layers, they left unaddressed whether such merging can work on downstream tasks. 
This work aims to close the gap and unify the findings of \cite{git-rebasin, zipit, merging_text_transformers} but for the speech and music domain. Different from the image and text domain, speech/music is a temporal signal with a high frame rate. We then simplify the pipeline so that the framework can be feasible in the speech/music domain. 

We test this correlation-based permutation approach on pre-trained models. In particular, it is noted that we test on models that were 1) initialized on different random seeds, 2) pre-trained with different pre-training data, and 3) optimized to learn different representation capabilities. Specifically, we focus on HuBERT and MERT as our case study, but this framework can be easily extended to any other models as long as the models to be merged share the same or very similar architecture, such as Whisper variants \cite{whisper}. We show that it is possible to merge a MERT encoder and a HuBERT encoder when we do channel-wise permutation of MERT towards HuBERT before performing linear interpolation. 

Results shown in Table \ref{tab:main_results} suggest our method can retain the speech performance of HuBERT while also improving downstream task performance on music tasks. The best performing model is achieved when the channels of the CNN encoder are permuted as well as the attention heads ordering and the channels within each attention head (ID 5 in Table \ref{tab:main_results}). This setup increases the average SUPERB score by 14.83 points. Additionally, ablations on which parts of a model are more suitable to permute are also provided with their respective statistical significance analysis.

\begin{figure*}[t!]
    \centering
    \includegraphics[width=\textwidth]{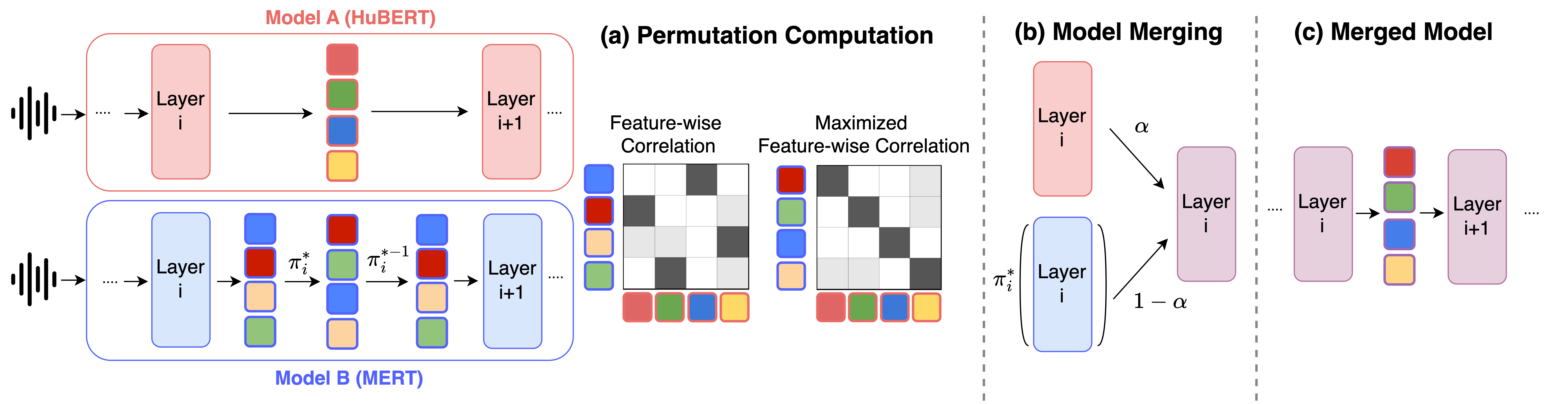} 
    \caption{\textbf{Feature‐wise Permutation Framework for Merging HuBERT and MERT.} 
    \textbf{(a) Permutation Computation:} For each layer $i$, compute a permutation $\pi_i^*$ that maximizes channel‐wise cross‐correlation between MERT’s feature activations (blue) and HuBERT’s feature activations (red). 
    The toy $4\times4$ correlation matrices illustrate raw correlations (Channel‐wise) on the left and reordered, near‐diagonal correlations (Maximized Channel‐wise) on the right. 
    \textbf{(b) Model Merging:} For all layer $i\in \mathcal{S}$, apply $\pi_i^*$ to reorder MERT’s weight matrix for layer $i$ and linearly interpolate the permuted MERT weights with HuBERT’s weights, producing a merged layer. \textbf{(c) Merged Model:} The resulting merged encoder (purple) incorporates speech representations from HuBERT and music representations from MERT into a single unified model.}
    \label{fig:permutation_framework_main}
\end{figure*}

\section{Related Work}

Early automatic-speech-recognition (ASR) research showed that combining multiple Hidden Markov Model (HMM) based acoustic models improves robustness. The work \cite{stolcke1994best} proposed to combine observation likelihoods and state transition probabilities. A Bayesian HMM-state merging is proposed by treating the problem via Bayesian model inference to decide which pair of states is best to merge at each step. Similarly, \cite{state_merging_dialectal_chinese} performs merging of HMM tied-states of a standard triphone-based HMM acoustic model to adapt from standard Chinese to specific dialects. Both works present a way of merging before the usage of neural networks in speech. Their focus is on better modeling of the observation likelihood of HMMs, which, back then, was modeled via Gaussian Mixture Models. While not the same as model merging, there have been several works towards model combination, one of many examples is ROVER \cite{rover}, which first aligns the word sequences generated by several ASR recognizers into a word-transition network. It then selects the consensus word at each network node via majority voting. Similar ideas have been explored for lattice and n-best hypothesis combination \cite{gales-lattice-comb,wong2020combination}, as well as for hybrid-models combination \cite{parikh-etal-2024-ensembles, schwenk00_icslp} and neural network models' combination \cite{hansen1990neural,checkpoint_ensembles}.

Unlike ensemble methods that combine predictions (e.g., by majority vote) \cite{hansen1990neural,checkpoint_ensembles}, model merging integrates multiple models into one via some forms of parameter averaging. This has the benefit that only one model is needed when data is passed through to make a prediction on a task.

One challenge in model merging is that independently trained networks often lie in different weight spaces or loss basins. When two models share a common initialization and are subsequently fine-tuned on different tasks, their final weights tend to remain in nearby regions of the loss landscape. As a result, linear interpolation between such fine-tuned models produces a new model that maintains good performance on each task \cite{what-transfer-learning}. An example of this approach is the task arithmetic framework \cite{task_arithmetic, tiesmerging, dare}. In task arithmetic, a base model is first fine-tuned on individual tasks; then each task vector is defined by subtracting the base model’s weights from the fine-tuned model’s weights. Combining knowledge across multiple tasks becomes equivalent to interpolating these task vectors. This method has been applied successfully in speech domains such as ASR, text-to-speech, and speech translation \cite{task_arithmetic_synthetic_speech,task_arithmetic_tts,task_arithmetic_speech_translation,selective_attention_merging_asr,ritter_distill_speech_music}. However, the requirement of a shared initialization limits its use when merging models trained on entirely different objectives or datasets.

Recent work has introduced permutation-based alignment to remove the shared-initialization constraint. Git Re-Basin \cite{git-rebasin} demonstrates theoretically that independently trained CNNs and MLPs often converge to permutation-equivalent minima. By finding a channel-wise permutation that aligns one network’s parameters with another’s, linear interpolation can traverse a low-loss path between the two, enabling merging without retraining. Building on that idea, ZipIt \cite{zipit} shows in practice that CNNs trained on disjoint tasks can be merged via permutations and still perform well on both tasks. Extensions of these permutation methods have begun to tackle transformer architectures. For instance, \cite{merging_text_transformers} describes how to permute components such as multi-head attention, focusing on loss barrier analysis for models that differ only by initialization seed. Because that work does not evaluate downstream task performance or handle models trained on different data and objectives, it leaves open the question of whether transformers can be merged under those conditions. The present study addresses this gap by applying channel-wise permutation alignment to both CNN and transformer layers, demonstrating the feasibility of merging transformers trained with different objectives and datasets.

\section{Channel-wise Permutation for Model Merging}
\label{sec:permute}

\subsection{Problem formulation}

Let $\textbf{f}_{{\theta}^{A}}$ and $\textbf{f}_{{\theta}^{B}}$ denote two independently-trained encoders that share the same backbone architecture (here, HuBERT and MERT, respectively) but were optimized on disjoint objectives and data. Here, $\theta^{A}$ and $\theta^{B}$ represent the model weights of models A and B.
Although their weight tensors have identical shapes, direct interpolation $\lambda\theta^{A}+(1-\lambda)\theta^{B}$, $ 0 \leq \lambda \leq 1$, is sub-optimal because each model has converged to a different weight space, such phenomenon has been studied in \cite{git-rebasin}. Consequently, we posit that for a more meaningful merging, it is required to discover a permutation $\pi^*$ such that the permuted parameters $\pi^*(\theta^{B})$ land in the same basin as $\theta^{A}$, after which linear interpolation safely traces a low-loss path between the models \cite{git-rebasin,zipit}. Formally, we seek
\begin{equation}
\pi^{\star}
    = \underset{\pi\in\mathcal{P}}{\arg\max}
      \sum_{l\in\mathcal{S}}
      \frac{
        \mathbb{E}\!\left[
          \left(\theta^{\text{A}}_{l} - \mu_{\theta^{\text{A}}_{l}}\right)^{\mathsf{T}}
          \left(\pi_l\!\left(\theta^{\text{B}}_{l}\right) - \mu_{\theta^{\text{B}}_{l}}\right)
        \right]
      }{
        \sigma_{\theta^{\text{A}}_{l}}\,
        \sigma_{\theta^{\text{B}}_{l}}
      }
\label{eq:permute_obj}
\end{equation}
where $\pi^{\star} = \{\pi_l^{\star}, \, \,\forall \,\, l\in\mathcal{S}\}$ be the set of optimal per-layer permutations, $\mathbb{E}$ denotes the expectation, $\mathcal{P}$ is the space of valid channel permutations, $\mu_{\theta^{\text{A}}_{l}},\mu_{\theta^{\text{B}}_{l}}$ is the mean across the feature dimension for model A and B, $ \sigma_{\theta^{\text{A}}_{l}}$, $ \sigma_{\theta^{\text{B}}_{l}}$ their standard deviations and $\mathcal{S}$ is a set of pre-defined feature activations where channel (feature) alignment is desired. 

Eq.~\eqref{eq:permute_obj} mirrors the \textit{Git Re-Basin} objective \cite{git-rebasin} but is solved per-layer rather than jointly. Additionally, Eq.~\eqref{eq:permute_obj} is cross-correlation across feature activations rather than $l_2$ distance.

In practice, handling Eq.~\eqref{eq:permute_obj} requires considering how to permute for sequential data like audio. Unlike typical applications in the CV domain, as in Git Re-Basin \cite{git-rebasin}, where temporal information is not present, our method involves a specific treatment of the feature activations. For each pre-defined activation point $l \in \mathcal{S}$, we reshape the feature to enable computation of correlations only across the features (channels), treating temporal and batch instances as samples over which channel statistics are aggregated. The objective then becomes finding the best permutation matrix that maximizes the new channel ordering of model B relative to the channel ordering of model A, as illustrated in Fig. \ref{fig:permutation_framework_main}. The best permutation $\pi^{\star}$ can be solved by the Jonker-Volgenant algorithm \cite{jonker_algorithm}. It is important to note that, as Eq.~\eqref{eq:permute_obj} is solved per layer, an un-permute operation is required so that the input data can pass through the layers as originally intended, this operation is important to maintain original functionality of the model so that at each layer we compute the correct $\pi_l^{\star}$. In practice, the inverse of the permutation $\pi_l^{\star}$ is used. An illustration can be seen in Fig. \ref{fig:permutation_framework_main} (a).

\subsection{MERT alignment towards HuBERT - CNN Layers}

When merging MERT towards HuBERT, it has to be decided which components are important for permutation alignment, i.e, decide the set $\mathcal{S}$. HuBERT and MERT share the same architecture. Namely, a convolutional encoder and a transformer encoder. In both models, the first CNN layer is composed of a Conv1D layer, Group Normalization \cite{group_norm}, and a GeLU activation \cite{gelu}, while the rest of the CNN layers do not include Group Normalization. We then compute channel-wise correlation alignment after the Group Normalization on the first layer and after the Conv1D for the rest of the layers.

In order to compute feature activations for cross-correlation analysis between HuBERT and MERT features as needed for Eq.~\eqref{eq:permute_obj}, a calibration dataset is needed. Then, at each $l \in S$ in the CNN encoder, we collect the corresponding feature activations of both models. For a batch of $B$ input audio segments, these activations have a shape of $\mathbb{R}^{B\times C_{l}\times T_{l}}$, where $C_{l}$ is the number of channels and $T_{l}$ the temporal dimension at node $l$. 

To enable an effective alignment of channels based on their representational similarities, we reshape these tensors into a two-dimensional form $\mathbb{R}^{C_{l}\times B \cdot T_{l}}$. This reshaping explicitly consolidates all non-channel dimensions (batch and temporal) into a unified dimension, thereby simplifying the correlation computation by directly exposing channel-wise feature statistics. Such a transformation is justified by the inherent nature of CNNs, whose learned weights are explicitly designed to be invariant across temporal positions. Hence, aggregating temporal and batch dimensions preserves the intrinsic semantics captured by each channel, allowing direct, meaningful comparison of channel-level features independent of their specific temporal or sample-wise locations. With these reshaped activations, we optimize the channel ordering of MERT towards HuBERT following Eq.~\eqref{eq:permute_obj}.

\subsection{MERT alignment towards HuBERT - Transformer Layers}
\label{subsec:transformer_perm}
This section explains assumptions taken and the components we decided to permute inside the transformer layers of model B (MERT) towards model A (HuBERT). As noted in \cite{merging_text_transformers}, merging transformer layers is complicated due to more complex connections than simpler MLP-like architectures. Our decisions are primarily motivated by \cite{merging_text_transformers}. 

The attention within a transformer encoder layer is typically composed of a query $Q\in \mathbb{R}^{T,B,D}$, key $K \in \mathbb{R}^{T,B,D}$ and value $V \in \mathbb{R}^{T,B,D} $ matrix, here $B$ denotes the batch size. Then, the attention is computed for multiple heads $h$ as follows,
\begin{align}
\label{eq:transformer_attn}
\text{Attn}(Q,K,V) &= \operatorname{softmax}\!\Bigl(\tfrac{QK^{\top}}{\sqrt{d_k}}\Bigr)V, \\[6pt]
\label{eq:transformer_head}
\text{head}_i &= \text{Attn}\bigl(QW_i^{Q},\,KW_i^{K},\,VW_i^{V}\bigr), \\
\label{eq:transformer_multihead}
\text{MultiHead}(Q,K,V) &= \operatorname{Concat}\bigl(\text{head}_1,\dots,\text{head}_h\bigr)W^{O}
\end{align}

Where \(W_{i}^{Q},\,W_{i}^{K},\,W_{i}^{V}\!\in\!\mathbb{R}^{D\times d_{k}}\) are the per–head projection matrices (\(d_{k}=D/h\)), and \(W^{O}\!\in\!\mathbb{R}^{h\cdot d_{k}\times D}\) learns to aggregate the information of the heads and projects the features back to the model dimension \(D\).

After attention, it follows a position-wise feed-forward sub-layer, defined as,
\begin{equation}
\begin{aligned}
\operatorname{FFN}(x) &= \text{GeLU}\!\bigl(xW_{1}+b_{1}\bigr)W_{2}+b_{2},
\label{eq:transformer_fnn}
\end{aligned}
\end{equation}

With $W_{1} \in \mathbb{R}^{D\times D_{\text{ff}}}$, $W_{2}\in \mathbb{R}^{D_{\text{ff}}\times D}$,  \(D_{\text{ff}}\ \geq D \).


As seen in Eq.~\eqref{eq:transformer_attn}-\eqref{eq:transformer_fnn}, there are many points where feature activations could be computed for solving Eq.~\eqref{eq:permute_obj}. Nonetheless, this is not that trivial; for example, attempting to independently permute the outputs of the Query ($QW_i^Q$), Key ($QW_i^K$), and Value ($VW_i^V$) projection matrices for each head using separate permutation matrices would likely be detrimental. The semantic relationship between a specific query, its corresponding key, and its associated value within a head is fundamental to the attention mechanism's ability to compute dot-product scores to correctly weight the values.

Furthermore, even if one were to permute $Q_i, K_i, V_i$ using the same permutation matrix (for instance, derived from correlating the attention scores), a critical challenge remains: the ordering of the $h$ attention heads. Ignoring the head ordering and simply permuting the concatenated output of Eq.~\eqref{eq:transformer_multihead} block can lead to suboptimal merging. This is because different heads often specialize in distinct functions \cite{clark2019does,voita2019analyzing}. For example, if MERT's head 1 functionally corresponds to HuBERT's head 3, but we try to align MERT's head 1 channels with HuBERT's head 1 channels without reordering, we are forcing a mismatch between functional units. This can lead to destructive interference during averaging, as the merged head would average parameters from two functionally disparate components.

Therefore, a strategy that accounts for such challenges must involve a hierarchical permutation as follows:
\subsubsection{\textbf{Permuting Attention Heads}} We first determine an optimal permutation $\pi_{heads}^*$ of MERT's $h$ attention heads to align them functionally with HuBERT's heads. This is achieved by correlating features associated with entire heads (i.e., after individual head computations but before the $W^O$ projection) for both MERT and HuBERT. Using these activations, we compute a head-to-head alignment quality matrix. An entry (j,k) in this matrix quantifies the maximum possible feature-level correlation achievable if MERT's j-th head were specifically aligned with HuBERT's k-th head. 
\subsubsection{\textbf{Permuting Features within Aligned Heads}} Once heads are matched, we proceed to align the features within these matched heads.
\subsubsection{\textbf{Permuting FFN Neurons}} For the FFN block, only the second linear layer ($W_2$) is permuted. This decision is based on preliminary experiments where permuting the first FFN layer ($W_1$) showed no improvements.


\section{Experimental Setup}

\subsection{Model Merging via correlation-permutation}

To analyze the proposed correlation-permutation approach for speech/music model merging, we select MERT Base\footnote{\scriptsize https://huggingface.co/m-a-p/MERT-v0-public} \cite{mert} as the music encoder and HuBERT Base\footnote{\scriptsize https://huggingface.co/facebook/hubert-base-ls960} \cite{hubert} as the speech one. Here, MERT will be permuted towards HuBERT so that HuBERT can be enhanced with musical abilities. 

We first select a subset of speech and music data for the computation of feature cross-correlations. In particular, we randomly select 5,000 samples of Music4All \cite{music4all} and LibriSpeech-100 \cite{librispeech} datasets. Such datasets were chosen because they consist of the data used to pre-train MERT and HuBERT, respectively. This 10,000 subset of music and speech data is used to compute the best permutation matrix that maximizes the channel-cross correlation across pre-defined layers $l \in \mathcal{S}$ of MERT and HuBERT. We use the LinearSumAssignment function from scipy\footnote{\scriptsize https://encr.pw/LinearSumAssignment} as done in \cite{git-rebasin,merging_text_transformers}, which guarantees a unique optimal solution at $O(n^3)$ time. Once the best permutation is found in a layer, we then compute the inverse of its permutation matrix ($\pi^{*-1}$) and unpermute the activations, ensuring clean statistics in the next layer. We then save all permutation matrices for $\mathcal{S}$ and apply them during the merging stage. For merging, we linearly interpolate following the setup in \cite{ritter_distill_speech_music}, which shows that using combination weights of 0.9 for HuBERT and 0.1 for MERT yields reasonable combined performance. All the experiments of Table \ref{tab:main_results} have been interpolated under this weight combination, except ID 3 that uses uniform weight averaging.

We experimented by introducing permutations at different parts of the model to analyze the merging effect on downstream task performance. These configurations determine which parts of the MERT model are subject to permutation alignment with HuBERT before the linear interpolation. The specific configurations in Table \ref{tab:main_results} (rows 6-10), are:
\subsubsection{\textbf{CNN}} $\mathcal{S}$ only considers the CNN encoder. Permutation is applied only to the channels of the convolutional layers (
described in Sec. \ref{sec:permute}.B). No permutation is performed within the transformer blocks.

\subsubsection{\textbf{CNN + ``ff\_only"}} Apart from CNN, it computes permutation to $W_2$ in Eq.~\eqref{eq:transformer_fnn}.

\subsubsection{\textbf{CNN + ``fnn+attn" (Best Method)}}It combines the permutation of the CNN encoder layers with a targeted permutation strategy within the transformer blocks. Specifically, the following components are also permuted:
(i) the $h$ heads ordering of the transformer layer (Eq.~\eqref{eq:transformer_multihead}), as detailed in Section \ref{sec:permute} C). (ii) The channels within each reordered head (iii)  $W_2$ in Eq.~\eqref{eq:transformer_fnn}.

\subsubsection{\textbf{CNN + ``all''}}  In addition to the components explained before, \emph{every} learnable matrix $W^{Q},\,W^{K},\,W^{V}$ in the transformer layer is permuted.

\subsubsection{\textbf{``fnn+attn''}}  Only the transformer is permuted as described in Sec.~\ref{subsec:transformer_perm}. The CNN stays un-permuted.

\subsection{Downstream tasks evaluation}

In order to evaluate the ability of the merged model to perform on speech and music tasks, we select tasks comprising the SUPERB Benchmark \cite{superb} as well as the MARBLE Benchmark \cite{marble}. For the speech tasks, we select the Automatic Speech Recognition (ASR)\cite{librispeech}, Keyword Spotting (KS)\cite{speech_commands}, Intent Classification (IC)\cite{fsc}, Speaker Identification (SID)\cite{voxceleb}, and Emotion Recognition (ER)\cite{iemocap} from SUPERB. For the music tasks, we select the Singer Identification (SingID)\cite{vocalset}, Vocal Technique Detection (VocID)\cite{vocalset}, Instrument classification (InstCls)\cite{nsynth}, Pitch classification (PitchID)\cite{nsynth}, and Genre classification (GenreID), in particular the multi-class classification task of Genre classification with the GTZAN dataset \cite{gtzan}. Additionally, we include the ESC50 task for environmental sound classification \cite{esc50}. All tasks follow the same configuration reported in \cite{ritter_distill_speech_music}. In particular, SUPERB tasks use default downstream model configurations, while the MARBLE tasks are implemented by adding 2 linear layers on top of the encoder representations. All the tasks learn a linear combination of the transformer layers on top of their defined downstream architecture. McNemar's test is conducted at a p-value of 0.05 to verify statistical significance. For 5-fold cross-validated tasks (ER, ESC50), contingency table cell counts were first summed across all folds, and a single McNemar's test was then performed on this aggregated table to provide an overall significance test. All significance tests were done to analyze if the best performing model performance (ID 5/8) is statistically significant against the linear interpolation and the ablations method CNN, CNN + ``ff\_only" and CNN + ``all'.

Additionally, to quantify average speech and music performance, we compute a speech score, a music score, and an overall average score. This scoring methodology adapts Eq. (1) from the ML-SUPERB benchmark \cite{mlsuperb}. In our application of this score, a key modification in this paper is the definition of the upper performance bound: instead of the general SOTA (State-Of-The-Art) value used in \cite{mlsuperb}, we employ $s_{t}(\text{BEST})$, which represents the best performance achieved on the primary metric for task $t$ by either the original HuBERT or MERT model. The score for a system $u$ is thus calculated as:
\begin{equation}
\label{eq:superb_score_custom}
\operatorname{Score}_u=\frac{1000}{|T|} \sum_t^T \frac{s_{t}(u)-s_{t}(\text {FBANK})}{s_{t}(\text {BEST})-s_{t}(\text {FBANK})}
\end{equation}
Here, $s_{t}(u)$ denotes the performance of the evaluated system $u$ on the metric for task $t$. Correspondingly, $s_{t}(\text{FBANK})$ is the performance of a baseline system using traditional FBANK features, and $s_{t}(\text{BEST})$ is the best performance between HuBERT and MERT on the task. $|T|$ represents the total number of tasks considered for the specific score (e.g., speech tasks for the speech score, music tasks for the music score).

This scoring framework is suited for evaluating our model merging experiments. By normalizing against $s_{t}(\text{BEST})$ the score directly reflects how well our merged model preserves or enhances the capabilities of the individual specialist models. It allows us to precisely measure the efficacy of the correlation-permutation approach in retaining HuBERT's speech performance as well as MERT's music abilities. Speech and Music score assesses representation ability across their own domains, and the higher the value, the better. Speech score accounts for the tasks: ASR, KS, IC, SID, and ER, while music scores for SingID, VocID, InstCls, PitchID, GenreID, and ESC50.

\section{Experiments}
\subsection{Baselines and  Linear Interpolation}

\begin{table*}[t!]
\setlength\tabcolsep{0.9pt}
\renewcommand{\arraystretch}{0.8} 
\centering
\scriptsize
\caption{HuBERT, MERT, and Merged Model Performance. $^{\dagger}$ denotes a statistically significant difference with the McNemar's test against the proposed model ID 5 at a p-value of 0.05.}
\label{tab:main_results}
\resizebox{\textwidth}{!}{%

\begin{tabular}{lccccccccccc>{\columncolor{lightgray}}c>{\columncolor{lightgray}}c>{\columncolor{lightgray}}c}
\toprule
\textbf{Upstream} & \textbf{ASR} & \textbf{KS} & \textbf{IC} & \textbf{SID} &\textbf{ER} & \textbf{SingID} & \textbf{VocID} & \textbf{InstCls} &\textbf{PitchID} & \textbf{GenreID} &\textbf{ESC50} & \cellcolor{headergray}\textbf{Speech} & \cellcolor{headergray}\textbf{Music} &\cellcolor{headergray}\textbf{Avg} \\
& \textbf{(WER\% $\downarrow$)} & \textbf{(Acc\% $\uparrow$)} & \textbf{(Acc\% $\uparrow$)} & \textbf{(Acc\% $\uparrow$)} & \textbf{(Acc\% $\uparrow$)} & \textbf{(Acc\% $\uparrow$)} & \textbf{(Acc\% $\uparrow$)} & \textbf{(Acc\% $\uparrow$)} &\textbf{(Acc\% $\uparrow$)} & \textbf{(Acc\% $\uparrow$)} &\textbf{(Acc\% $\uparrow$)} & \cellcolor{headergray}\textbf{(Score $\uparrow$)} & \cellcolor{headergray}\textbf{(Score $\uparrow$)} &\cellcolor{headergray}\textbf{(Score $\uparrow$)} \\
\midrule
\multicolumn{3}{l}{Main Results / Baselines}\\
\midrule
1) HuBERT (HB)     & 7.84  & 96.30 & 98.34 & 81.42 & 64.92 & 70.25 & 61.32 & 62.74 & 70.04 & 63.45 & 70.20 & 1000   & 885.63 & 942.82\\
2) MERT (MR)       & 23.61 & 89.45 & 59.21 & 29.49 & 55.19 & 79.00 & 72.54 & 64.97 & 91.26 & 70.69 & 73.97 & 665.53 & 1000   & 832.77 {\tiny \textcolor{red}{($-11.7\%$)}} \\
3) (HB+MR)/2       & 37.44 & 57.06$^{\dagger}$ & 16.77$^{\dagger}$ & 5.87$^{\dagger}$  & 52.20$^{\dagger}$ & 52.17$^{\dagger}$ & 47.72$^{\dagger}$ & 43.58$^{\dagger}$ & 41.28$^{\dagger}$ & 35.61$^{\dagger}$ & 34.35$^{\dagger}$ & 385.19 & 567.92 & 476.55 {\tiny \textcolor{red}{($-49.5\%$)}}  \\
4) 0.9 HB + 0.1 MR & 9.47  & 95.23$^{\dagger}$ & 97.02$^{\dagger}$ & 77.24$^{\dagger}$ & 65.12$^{\dagger}$ & 71.33$^{\dagger}$ & 61.75$^{\dagger}$ & 62.04$^{\dagger}$ & 70.03$^{\dagger}$ & 69.56$^{\dagger}$ & 67.05$^{\dagger}$ & 981.80 & 894.42 & 938.11 {\tiny \textcolor{red}{($-0.5\%$)}} \\
5) Proposed      & 8.88  & 96.17 & 97.79 & 80.22 & 65.61 & 72.74 & 63.98 & 63.79 & 70.00 & 69.62 & 71.69 & 997.72 & 917.58 & \textbf{957.65}  {\tiny \textcolor{green}{($+1.4\%$)}} \\
\midrule
\multicolumn{7}{l}{Permutation Ablations - Position where features are permuted}\\ 
\midrule
6) CNN & 9.52 & 95.43$^{\dagger}$  & 94.06$^{\dagger}$ & 77.44$^{\dagger}$ & 62.95$^{\dagger}$ & 73.31 & 59.30$^{\dagger}$ & 61.47$^{\dagger}$ & 70.22$^{\dagger}$ & 68.28$^{\dagger}$ & 66.07$^{\dagger}$ & 961.27	& 886.62 & 923.95\\
7) CNN + ``ff\_only" & 9.92 & 95.59	& 95.10	& 77.61$^{\dagger}$	& 66.64$^{\dagger}$	& 72.38	& 60.26	& 61.16$^{\dagger}$ & 70.53$^{\dagger}$ & 68.62$^{\dagger}$ & 65.45$^{\dagger}$ & 988.43 & 886.06 & 937.24\\
5) CNN + ``fnn+attn" & 8.88  & 96.17 & 97.79 & 80.22 & 65.61 & 72.74 & 63.98 & 63.79 & 70.00 & 69.62 & 71.69 & 997.72 & 917.58 & \textbf{957.65}   \\
8) CNN + ``all" & 9.70 & 95.62$^{\dagger}$ & 94.62$^{\dagger}$ & 77.70$^{\dagger}$ & 64.91$^{\dagger}$ & 71.39 & 59.91$^{\dagger}$ & 60.91$^{\dagger}$ & 70.23$^{\dagger}$ & 68.62$^{\dagger}$ & 64.80$^{\dagger}$ & 976.46 & 880.52 & 928.49\\
9) ``fnn+attn" &  9.08	& 95.42$^{\dagger}$	& 96.97$^{\dagger}$	& 75.97$^{\dagger}$	& 65.91	& 73.52	& 62.54	& 61.23$^{\dagger}$	& 71.44$^{\dagger}$	& 70.34	& 67.05$^{\dagger}$ & 985.30 & 903.22 & 944.26\\
\bottomrule
\end{tabular}
}
\vspace{-2mm}
\end{table*}

\begin{table}[t!]
\centering
\vspace{-2mm}
\caption{Percentage of Permuted Channels in MERT for ID 5 in Table \ref{tab:main_results}}
\label{tab:permuted_channels_summary}
\begin{tabular}{@{}lc@{}}
\toprule
Model Section                & Avg. \% Permuted Channels \\
\midrule
CNN Layer 0                  & 30.86\%                   \\
CNN Layers 1-6 (Avg.)        & 99.58\%                   \\
Transformer Attn Output (Avg.)  & 99.93\%                   \\
Transformer $W_2$ (Avg.)   & 99.97\%                   \\
\bottomrule
\end{tabular}
\end{table}

The performance of various merging strategies is detailed in Table \ref{tab:main_results} where it can be seen that uniform weight averaging of HuBERT and MERT (ID 3) leads to a degradation of performance across all the tasks, achieving a reduced speech score of 385.19 and a music score of 567.92, averaging a final score of 476.55 which is in relative percentage almost 50\% worse than HuBERT. 

Motivated by prior work \cite{task_arithmetic,ritter_distill_speech_music}, which suggests that linear interpolation benefits from asymmetric weighting (e.g., 0.9 and 0.1 rather than 0.5 and 0.5), we also evaluated a model with 0.9 on HuBERT weights and 0.1 on MERT weights (ID 4 in Table \ref{tab:main_results}). While this approach (average score: 938.11) is considerably better than uniform averaging, it still results in a slight 0.5\% performance decrease compared to the original HuBERT average score of 942.82. This reduction in performance suggests that even with weighted averaging, the models' distinct pre-training criterion, despite sharing the same architecture, lead to HuBERT and MERT likely occupying distinct regions in the weight space. 

\subsection{On the Effect of Component-wise Permutation for Merging}

To better understand the individual contributions of permuting different architectural components, we conducted an ablation study, with results also presented in Table \ref{tab:main_results}. The proposed method, ID 5, which involved permuting the CNN encoder layers and specific transformer sub-layers (``ffn+attn" setup), achieves the highest average score of 957.65. This configuration successfully improves the music score to 917.58 (from HuBERT's 885.63) while maintaining a strong speech score of 997.72 (HuBERT baseline: 1000). We performed statistical tests and found that all the results of ID 5 are significant when compared to the linear interpolation method ID 4.

We also further examined different permutation variants, which reveal certain insights, in particular:

\subsubsection{\textbf{Only CNN Permutation is Insufficient}} Permuting only the CNN layers (ID 6) yields an average score of 923.95. It can be noticed that permuting only the CNN component yields an inferior model compared to ID 4. This indicates that aligning the initial CNN extractor is not enough, as mismatches in the deeper transformer layers still significantly impact performance if unaddressed. It is interesting to notice that permuting the CNN layers seems to have a large effect on the SID  task, which needs to model information related to the speaker. Such a phenomenon may be explained by the fact that earlier layers in HuBERT are more important for SID, as studied in \cite{layerwiselivescu,livescu_all_models}, and then aligning CNN layers may avoid interference for modeling speaker information aspects. Contrarily, permuting only the transformer layers without the CNN part (ID 9) seems to affect SID.
\subsubsection{\textbf{Targeted Transformer Permutation is Highly Effective}} When only the targeted transformer components (``fnn+attn" ID 9, without CNN permutation) are permuted, the model achieves an average score of 944.26. This is notably higher than permuting CNN layers alone (923.95, ID 6) and highlights the significant impact of aligning these specific transformer sub-layers. In particular, it is noticeable the music score improvement (903.22). As stated before, CNN merging seems to have a high incidence on SID, and it can be noticed by the notorious SID performance drop. Additionally, it can be seen that permuting both CNN and ``ffn+attn" is complementary, achieving the most complete model in speech and music tasks.
\subsubsection{\textbf{Selectivity in Transformer Permutation}} The CNN + ``all” configuration (ID 8), where all query, key, values within the transformer blocks are permuted in addition to the ``ffn+attn” setup, results in an average score of 928.49. This is lower than the more selective CNN + 'fnn+attn' approach (957.65, ID 5) and even lower than fnn+attn alone (944.26, row 10). This suggests that attempting to permute every possible component within the transformer is not optimal and could be detrimental by disrupting more complex learned relationships or leading to less stable alignments.
The CNN + ``ff\_only" setting (ID 7), where CNN layers and only the first feed-forward (FFN) layer of the transformers are permuted, yields an average score of 937.24. While this improves upon permuting CNN layers alone (923.95, ID 6), it notably underperforms the CNN + ``fnn+attn" strategy (957.65, ID 5) and is slightly less effective than the non-permuted 0.9 HB + 0.1 MR baseline (938.11, ID 4). This outcome suggests that simply adding any transformer layer permutation is not guaranteed to be optimal, reinforcing the importance of targeting specific sub-components within the transformer architecture for alignment.

In summary, the ablation study shows that a carefully chosen permutation strategy targeting both the CNN encoder and specific components within the transformer blocks, namely the ordering of attention heads, the channels within these reordered heads, and the second linear layer ($W_2$) of the FFN module is essential for effectively merging independently trained HuBERT and MERT models. Simply permuting shallower layers or attempting to permute all transformer components indiscriminately leads to suboptimal outcomes.

\subsection{Layer-wise Permutation Analysis}

To understand how structural alignment varies across the depth of the models, we examined the percentage of channels permuted on each layer considered for permutation on MERT when using the CNN + ``fnn+atnn" setup. Results are shown in Table \ref{tab:permuted_channels_summary}, where interestingly, it can be seen that most layers are completely permuted with the exception of the first CNN layer of MERT, where only 30.86\% of channels were reordered. This suggests that the initial feature representations learned by both MERT and HuBERT at this shallow depth share considerable similarity. It is plausible that this first layer in both models learns to extract fundamental, low-level acoustic features, akin to filterbank-like representations. Therefore, the internal channel ordering for these basic features might already be substantially aligned between the two independently trained models, necessitating fewer permutations.

In contrast, all subsequent layers exhibit a much higher degree of permutation, with percentages consistently above 98.8\%. This indicates that as features become more abstract and specialized for the respective domains (speech for HuBERT, music for MERT) in deeper convolutional and transformer layers, their internal learned representations diverge significantly, requiring extensive reordering to find an optimal alignment.

\section{Conclusions}

This work introduced and validated a correlation-permutation methodology for merging independently pre-trained speech (HuBERT) and music (MERT) encoders that typically reside in disparate weight spaces. This yields an encoder that can generalizes better across all the evaluated tasks which could serve in applications beyond the evaluated in this paper.

Our primary contribution is a practical framework that enables the creation of a unified audio model without the previous works constraint of starting from a commonly initialized model. Such contribution is achieved by aligning the internal channel representations of the music encoder (MERT) with those of the speech encoder (HuBERT) on a layer by layer basis. This alignment is achieved by computing permutation matrices that maximize the features cross-correlations on their respective layers prior to performing linear interpolation of their parameters. Experiments have shown that the proposed alignment strategy allows the merged model to retain HuBERT's speech processing capabilities while significantly enhancing its performance in music tasks.

Ablation studies showed that the optimal merging performance is achieved by permuting the CNN encoder as well as specific components within the transformer blocks, the latter proved to be crucial for obtaining a good merged model.

Future work will extend the proposed approach to allow the merging of models that does not share the same architecture. While the current approach proves useful for merging MERT and HuBERT, a natural next step is to be able to combine models with greater degree of diversity on their representation capabilities. Additionally,  this work focused on merging two encoders. Scaling permutation-based merging to combine knowledge from more than two models remains an open question which will also be addressed in future work.

\newpage
\bibliographystyle{IEEEtran}
\bibliography{mybib}

\end{document}